\documentclass[12pt]{article}
\usepackage[hyperfootnotes=false]{hyperref}
  \usepackage{graphicx}
  \newlength{\abstractwidth}
  \renewcommand{\thefootnote}{\fnsymbol{footnote}}
  \renewcommand{\thanks}[1]{\footnote{#1}} 
  \newcommand{\starttext}{
  \setcounter{footnote}{0}
  \renewcommand{\thefootnote}{\arabic{footnote}}}
  \renewcommand{\theequation}{\thesection.\arabic{equation}}
  \newcommand{\be}{\begin{equation}}
  \newcommand{\bea}{\begin{eqnarray}}
  \newcommand{\eea}{\end{eqnarray}}
  \newcommand{\beq}{\begin{equation}}
  \newcommand{\ee}{\end{equation}}
  \newcommand{\eeq}{\end{equation}}

  \def\ba{\begin{eqnarray}}
  \def\ea{\end{eqnarray}}

  \def\12{{1 \over 2}}
  \def\eq{&=&}

  \def\cc{cosmological constant }
  
  \def\simleq{\; \raise0.3ex\hbox{$<$\kern-0.75em
      \raise-1.1ex\hbox{$\sim$}}\; }
   \def\simgeq{\; \raise0.3ex\hbox{$>$\kern-0.75em
      \raise-1.1ex\hbox{$\sim$}}\; }

\def\cdl{Coleman De Luccia}

\def\ba{\bf{a}}

  \def\h3{{\cal{H}}_3}

\begin{document}
  \renewcommand{\theequation}{\thesection.\arabic{equation}}
  \begin{titlepage}
  \rightline{SU-ITP-10/34}
  \bigskip

  \bigskip\bigskip\bigskip\bigskip

    \centerline{\Large \bf {Crunches, Hats, and a Conjecture}}

  \bigskip \bigskip

  \bigskip\bigskip
  \bigskip\bigskip

\begin{center}
  {{\large Daniel Harlow and Leonard Susskind  }}
  \bigskip

\bigskip
Stanford Institute for Theoretical Physics and  Department of Physics, Stanford University\\
Stanford, CA 94305-4060, USA \\

\vspace{2cm}
  \end{center}

  \bigskip\bigskip
  \begin{abstract}

Our purpose in this paper is to discuss criteria for the existence of a precise dual description of a cosmology. A number of exact descriptions exist for flat and anti de Sitter backgrounds and possibly for  open FRW universes that nucleate in an eternally inflating background. In addition duals have been proposed for de Sitter space, and for crunching FRW bubbles with negative cosmological constant. In the latter cases there is reason to think the dualities are at best approximate. One of our primary purposes is to analyze the quality of these descriptions, i.e., how exact they can be made. Maldacena's recent discussion of dualities involving crunching FRW cosmologies provides an opportunity for exploring some of these question.

  \medskip
  \noindent
  \end{abstract}

  \end{titlepage}
  \starttext \baselineskip=17.63pt \setcounter{footnote}{0}


\bigskip
\setcounter{equation}{0}
\section{Six Steps to a Crunch}

In this paper we will be interested in the question of how precise a  quantum description of a given spacetime is possible. Examples include the S-matrix description of flat space, Matrix Theory descriptions of various backgrounds, the ADS/CFT duality, various conjectured dual descriptions of de Sitter space, FRW/CFT duality, and recent dual descriptions of crunching ADS cosmologies. The question we wish to raise is whether such descriptions can or should have more mathematical precision than the maximum precision of observations within these geometries. We suggest that the answer is no; the mathematical description of a spacetime should have as much ambiguity as the most precise observations.

To illustrate, consider the case of de Sitter space. Global observations on scales larger than the horizon are impossible since information cannot be gathered and correlated from outside the horizon. For observations within the horizon, we are limited by the finite maximum entropy inside a causal patch. The numerical accuracy of any real measurement is constrained to be no better than what can be expressed using a maximum number of binary digits.

These limitations derive from two quantum-mechanical facts. The first is that the entropy of a finite quantum system is finite. This is not true in classical physics since classical phase space is continuous. To define the entropy of a classical system a cutoff or coarse graining must be invoked. Without the cutoff, an arbitrarily large amount of information can be stored in a system with arbitrarily small energy. For example, just specifying the energy to $10^{1000} $ binary decimal places determines that many bits of information.
Quantum mechanics provides the phase-space cutoff that makes entropy and information finite.

The second fact is that observers and apparatuses are necessarily part of the system. One cannot keep invoking bigger systems that observe from the outside. These two facts together limit the precision of an observation.
The only hope for an infinitely precise description is that the universe is unbounded in degrees of freedom.

Note that the imprecision we are talking about has nothing to do with the lack of knowledge of details. The physics of a black hole will always have a degree of ambiguity unless we know the details of the initial state that formed the black hole. But in principle we can know those details.

Exact descriptions do exist for describing certain  background space-times. Physics in flat spacetime is believed to have an exact unitary S-matrix and in some cases the flat backgrounds can be described by Matrix Theory. Anti de Sitter spaces have dual CFT descriptions which are believed to be exact. Lastly, open FRW bubbles nucleated in an eternally inflating space may have a description through the FRW/CFT duality \cite{hat}, although this last case not nearly as certain as the others.

By contrast eternal de Sitter space does not seem to have an exact dual \cite{Dyson:2002nt} \cite{Goheer:2002vf}
That does not mean that dS/CFT \cite{ds-cft}, or dS/dS \cite{dsds} may not be  useful approximations.

A case that has been studied recently \cite{juan} is open FRW universes with negative \cc \ that eventually become singular  (ADS-crunches). Crunches of this type are especially interesting since they, along with hat-geometries \cite{hat} constitute the terminal endpoints of eternal inflation. In other words the future conformal boundary of eternal inflation is composed of crunches and hats plus an inflating fractal of measure zero.

As we will see, there are some cases of  ADS-crunches in which the duality can potentially be exact,  but in most cases, particularly when the FRW universe nucleates from \cdl \ tunneling during eternal inflation, the duality proposed in \cite{juan} is at best inexact.

It would be  very interesting have a deep criterion for when a  description can be made exact and when it cannot. One may go further and ask if a figure-of-merit can be defined that quantifies the unavoidable degree of ambiguity in describing a given cosmology quantum mechanically.


We will begin by reviewing Maldacena's proposal for dual descriptions of crunching ADS geometries. There are six steps to the  construction, five of which were spelled out in  \cite{juan}.

\subsection{The Steps}
 \begin{figure}[h!]
\begin{center}
\includegraphics[width=18cm]{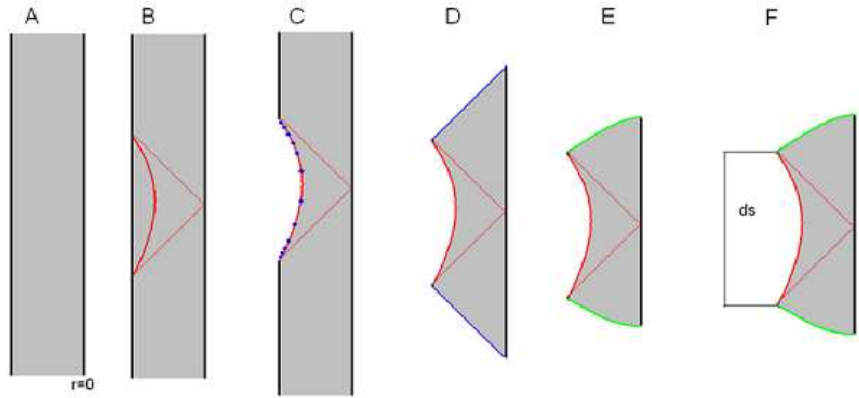}
\caption{The six steps in defining a dual to a crunch.
} \label{1}
\end{center}
\end{figure}

\begin{itemize}
\item A.
Begin with an ADS geometry as in Fig 1A.
\item B.
Slice the geometry by hyperbolic slices as in Fig 1B. One such slice will be replaced by a domain wall or end-of-the-world brane. Call that slice D.
    \item C.
    Integrate out the degrees of freedom between D and the boundary of ADS. Following Heemskerk and Polchinski \cite{joe}, this corresponds to integrating out degrees of freedom in the boundary CFT and leaves a cutoff field theory on D. The cutoff length scale is the ADS radius $l_{ads} $ \cite{susskind witten}. In Fig 1.C the region to the left of D has been excised. The surface D has been shown broken up into cutoff sized segments by the blue dots.
        \item D.
        So far we have not really changed the physics in the grey region. We have just given a renormalized Wilsonian description of it. But now we add a perturbation on D consisting of an irrelevant operator. According to Maldacena such a perturbation describes a source of particles that propagate into the ADS region. The symmetry of the problem is that of an open FRW universe and the particles are homogeneously distributed. The result is well known: the geometry crunches.

            For very small values of the source the singularity will be very close to light-like as shown by the blue line in Fig 1D.
            \item E.
            For finite irrelevant perturbation the singularity is space-like as in Fig 1E. This is just the ordinary crunch in open FRW with negative \cc \ and finite energy density.

                \end{itemize}

                Maldacena argues that this set of steps implies a duality between crunching FRW geometries and cutoff quantum field theory on 2+1 dimensional de Sitter space. One concern about this duality is that,  the cutoff field theory will in general not be ultraviolet completable. (There are some exceptions that we will discuss.) The UV incompleteness is an especially  serious problem for Minkowski space quantum field theories, and even more problematic in de Sitter space.
                One has two choices: break lorentz invariance as in Hamiltonian lattice gauge theory, or break unitarity by regulating in Euclidian signature and analytically continuing.
                The problem is especially acute in de Sitter space where the number of degrees of freedom grows exponentially with time. How those degrees of freedom are to be added is not clear. Ultraviolet incompleteness is the main source of ambiguity on the dual side.

                Before going on to the last step, we will discuss a strategy for constructing a UV complete version of steps A through E\footnote{The model described below was suggested to us by Eva Silverstein.}. Consider ADS/CFT on a four dimensional bulk Euclidean ADS space. The boundary dual is a CFT on a 3-sphere. Now perturb the boundary theory with a relevant operator  so that it flows to a second infrared fixed point CFT'. If this is done on the 3-sphere the bulk geometry will be spherically symmetric. The outer region will be ADS and as the radial variable decreases, the geometry will make a transition to ADS'. The transition region plays the role of the (Euclidean) domain wall. From the  point of view of the dual field theory, the infrared behavior is described by CFT' plus some irrelevant operators.

                Continuing the bulk geometry to Minkowski metric, the spherical Euclidean domain wall becomes the de Sitter wall D. Maldacena's arguments should apply and lead to a crunching FRW geometry. Examples of this type appear to be  UV complete versions of the duality between QFT on de Sitter space and ADS-crunches.

                In other cases it typically will not be possible to complete the cutoff version of the boundary field theory on D to a continuum field theory. If there is no UV fixed point then running the theory to small distances will inevitably hit a wall at which irrelevant couplings blow up. This is the phenomena of a Landau pole. In these cases the theory is only approximate and lacks precision.

                UV complete examples are interesting but  they are not what we want for bubbles that nucleate in an eternally inflating vacuum. That brings us to step F.

              \begin{itemize}
               \item F. The final step, not discussed in \cite{juan},
               is to attach the geometry to a portion of de Sitter space as in Fig 1F.

              \end{itemize}

This last step adds an altogether new level of ambiguity.
The  hyperboloid D is 2+1 dimensional de Sitter space (not to be confused with the 3+1 de Sitter ancestor). Using arguments similar to those in \cite{dsds} one can show that the theory on D cannot just be a cutoff quantum field theory, but it must contain 2+1 dimensional quantized gravitational degrees of freedom. In view of the fact that there is no precise description of quantum gravity in de Sitter space it is clear that step F leads to an ambiguous theory.

The arguments for dynamical gravity on D are not new and are similar to those found in \cite{Randall:1999ee} and \cite{dsds}. Although these arguments are convincing, we will verify them by explicitly computing fluctuations of the metric on D and showing that they are consistent with the presence of a dynamical gravitational theory on the domain wall. This will allow us to make contact with the FRW/CFT proposal.

\setcounter{equation}{0}
\section{FSSY Analysis}

\subsection{The Coleman DeLuccia Instanton}
There are of course  fluctuations of the intrinsic geometry of D due to the ordinary 3+1 dimensional gravitational fluctuations in the bulk. The same could be said about the ordinary slicing of ADS by a cutoff surface. However, because ADS is asymptotically cold, those fluctuations do not require a gravitational sector on the cutoff surface. A good diagnostic for whether D is asymptotically warm--that is whether gravitational degrees of freedom are required on D--is to consider the infrared behavior of the induced metric fluctuations far out on the asymptotic future of D. In other words, thinking of D as 2+1 de Sitter space, does the metric at future infinity have the characteristic logarithmic fluctuations of a dynamical gravity theory of de Sitter space? To answer this question we consider two-point correlators of certain components of the induced metric on D. In particular we consider components of the metric in the surface D.

The method of computation of the two-point function is the same as in \cite{fssy}.  The gravitational fluctuation satisfies the equation of motion of a minimally coupled, massless scalar field $\chi$, so we will mainly concentrate on the scalar case and refer to \cite{fssy} for the tensor case.  We will compute the correlator in the Euclidean version of the bubble geometry, and then continue back to Lorentzian signature.  The relevant geometry is the \cdl \ instanton, \cite{Coleman} portrayed in Figure 2.
The symmetry of the instanton ensures the metric has the form:
\be
ds^2 = dx^2 + a(x)^2 d\Omega_3^2
\label{2.1}
\ee
where
\be
d\Omega_3^2  = d\theta^2+\sin^2 \theta \ d\Omega_2^2
\ee
$x$ is a Euclidean time variable.  $a(x)$ is determined by solving the Euclidean Einstein-matter equations of whatever matter fields are involved in the decay, for example a scalar field with two minima.

As in \cite{fssy} the coordinate $x$ may be replaced by a conformal coordinate $X$ defined by
 \be
 dX= dx/a(x).
 \ee
 The variable $x$ runs over a finite interval from $0$ to $x_0$ but the conformal time runs from $-\infty$ to $+\infty$. The metric becomes
 \be
 ds^2=a(X)^2 (dX^2 + d\theta^2 + \sin^2 \theta \ d\Omega_2^2)
 \ee
 The asymptotic behavior of $a(X)$ follows from the smoothness of the instanton at the poles, $x=0$ and $x=x_0.$ One finds
 $$
 a(X) \to e^{-|X|}
 $$
 for large $|X|.$

In the thin-wall approximation the instanton is constructed by gluing a portion of a 4-sphere to a portion of the 4-dimensional hyperbolic plane. The spherical region is the Euclidean continuation of de Sitter space, and the hyperbolic region is the continuation of ADS. The surface where they meet is the continuation of the domain wall D.  Let us use the notation $l_{ads}$  and $l_{ds}$ to denote the radii of curvature of the spherical and hyperbolic regions. In the the spherical region $a(x)$ has the form
\be
a(x) = l_{ds}\sin{\left[ l_{ds}^{-1} \ (x_0 -x) \right] },
\label{2.3}
\ee

while in the hyperbolic region
\be
a(x) = l_{ads}\sinh{ [l_{ads}^{-1} \ x]  }
\label{2.4}
\ee

 \bigskip
 \begin{figure}
\begin{center}
\includegraphics[width=10cm]{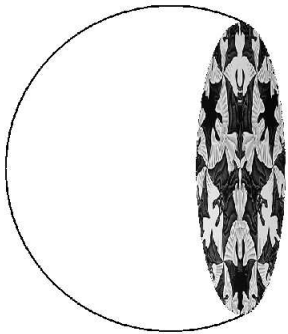}
\caption{Coleman DeLuccia instanton for a transition from de Sitter space to a vacuum with negative \cc. The spherical region to the left is the de Sitter region. The region to the right is a piece of a hyperbolic space with negative curvature. } \label{2}
\end{center}
\end{figure}
 \bigskip

 We can go to conformal Euclidean time, and we can choose the domain wall to be at X=0. We then have:\footnote{Note that here $\theta(X)$ is the heaviside theta function, not the angle $\theta$!}
\be
\label{metric}
a(X)=-\theta(-X) \frac{l_{ads}}{\sinh(X-X_0)}+\theta(X)
\frac{l_{ds}}{\cosh(X+X_0')}
\ee
Here $X_0$ is positive to avoid a singularity, and is related to $X_0'$ by continuity of the metric at $X=0$.  They both can be determined from the domain wall tension $\sigma$, the formula is:
\be
\label{tension}
\alpha\equiv4\pi G \sigma a(0)=\coth(X_0)+\tanh(X_0')
\ee
This can be derived from equation (3) in \cite{daniel}.  One can also show that a solution always exists for any values of $\sigma$, $l_{ds}$, and $l_{ads}$.  It should be emphasized that the thin-wall approximation here is an example rather than a crutch, as the central conclusion of the non-decoupling of gravity will be true for general CdL bubbles.

\subsection{The Bound State}
 To calculate the correlation function of $\chi$ we will pick two points, $1$ and $2$. With no loss of generality we may pick $\theta_1 =0$ and simply call $\theta_2 = \theta$. The symmetry ensures no dependence on the other angular variables.  Following \cite{fssy} we define the correlation function
 \be
 \hat{G}(X_1,X_2;\theta) = a(X_1)a(X_2)\langle   \chi(X_1,0)\chi(X_2,\theta)
 \rangle
 \ee

 The minimally coupled scalar field equation leads to the following Schrodinger-like (inhomogeneous) equation:\footnote{In this equation $\delta(\theta)$ is not really well defined, since we have placed the delta function at a place where the coordinates break down.  What we really mean is that the correlator obeys the wave equation away from $\theta=0$, and that near $\theta=0$ the correlator scales like $1/\theta$.  Moreover, as explained in \cite{fssy}, the shift symmetry of the field means that the zero mode can't obey the wave equation unless we introduce a background charge.  Both these technicalities are taken into account in the definitions of the angular wave functions below.}
 \be
 \left[
 -\partial^2_{X_1} + U(X_1) - \nabla^2\right]  \hat{G}(X_1,X_2;\theta)
 =\delta(X_1 -X_2){\delta(\theta)\over \sin^2{\theta}}
 \ee
 where $\nabla^2$ is the Laplacian on $\Omega_3.$ The potential $U$ is given by
 \be
 U(X)= {1 \over a(X)}{d^2a(X) \over dX^2  } = {a'' \over a}
 \ee

 The method of calculating $\hat{G}$ is to expand it in eigenmodes of $[-\partial^2_{X_1} + U(X_1)].$ Thus we consider the ordinary Schrodinger equation
 \be
[ -\partial^2_{X_1} + U(X_1)]u(X)= \lambda u(X)
 \ee

 This turns out to have a single normalizeable bound state, which has $\lambda=0$ and is given by:
 \be
 u_B(X) \propto a(X)
 \label{2.12}
 \ee
 The existence of this bound state follows from the general form of the CDL instanton, and has no assumptions about the thin-wall approximation in it.  That it is normalizeable follows from the boundary conditions of the CDL instanton.  That there are no other bound states follows from a supersymmetric quantum mechanics argument presented in \cite{fssy}.  Thus the rest of the spectrum is in the continuum. The eigenvalues are $\lambda=k^2+1$, and the wave functions $u_k(X)$ describe scattering off of a potential wall.

As in \cite{fssy}, the full correlator can now be written:
 \be \label{2point}
\hat{G}(X_1,X_2,\theta)=\int_{-\infty}^{\infty} \frac{dk}{2\pi}G_k(\theta)u_k(X_1) u_k^*(X_2)+u_B(X_1)u_B(X_2) G_B(\theta)
\ee
Here $G_k$ are angular wave functions, given by:
\be
G_k(\theta)=\frac{\sinh{k(\pi-\theta)}}{\sin\theta \sinh{\pi k}}
\ee
\be
G_B(\theta)=\frac{\cos\theta}{\sin\theta}\left(1-\frac{\theta}{\pi}\right)
\ee
These formulas are evaluated explicitly for the case of the thin-wall geometry in appendix A.

 \subsection{The Continuation}
The analytic continuation we are interested in here is simpler in the present case than in \cite{fssy}. That is because we are interested in correlation functions on the domain wall, and the domain wall is part of the Euclidean geometry. The only continuation that is necessary is
 \be
 \label{continuation}
 \theta \to i\omega +\frac{\pi}{2}
 \ee
 The significance of $\omega $ is simple.  Consider the ``waist" of the hyperboloid D. Then $\omega$ is the hyperbolic angle from the waist. In other words it is the dimensionless global time in the de Sitter geometry describing D.  Note that this continuation is identical for any $a(X)$.

 In \cite{fssy} the continuation was to the interior of the FRW patch, in which $\theta$ is continued to a space-like radial variable $R$.  The continuation of $\theta$ is again independent of $a(X)$, but one also has to continue $X$ to a timelike FRW conformal coordinate $T$.  The detailed form of this second continuation depends on $a(X)$.  The continuation for $\theta$ is:
\be
\theta \to iR
\ee
\be
\sinh[X-X_0] \to -\cosh[T]
\ee

Both $R$ and $\omega$ are hyperbolic angles and transform the same way under the $O(3,1)$ symmetry of the instanton.  These relations are shown in Figure 3.

  \bigskip
 \begin{figure}
\begin{center}
\includegraphics[width=12cm]{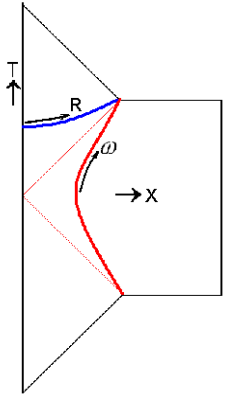}
\caption{The action of the O(3,1) group on the Coleman DeLuccia geometry.} \label{3}
\end{center}
\end{figure}
 \bigskip

To do this analytic continuation, we need to discuss the analytic properties of the expression (\ref{2point}) for the correlator as a function of $\theta$.  Without making any assumptions about $a(X)$ we do not know the analytic dependence of the wave functions on $k$.  We know that they must asymptote at large and negative $X$ to oscillating exponentials, but we will be interested in computing correlators in the vicinity of the domain wall.  So we will have to make a ``reasonable'' assumption that the wave functions don't grow exponentially with $k$ in the region where we are studying them.  This is certainly true for the thin-wall wave functions presented in the appendix, and it seems that convergence of the $k$ integral when one inserts a complete set of scattering states will ensure it in general.  In that case, the convergence of the integral in (\ref{2point}) is entirely dominated by the $k$ dependence of $G_k(\theta)$.  It is then easy to see that the integral will converge as long as $0<\Re(\theta)<2 \pi$.  Since our continuation (\ref{continuation}) gives a $\theta$ in this region, we may take the Lorentzian correlator to be given by the Euclidean expression (\ref{2point}) with the integrand appropriately continued but the contour left on the real $k$-axis.

We are interested however in extracting the fixed $X$, large $\omega$ behaviour of the correlator, and it turns out that we can deform the contour in a way that makes the asymptotics transparent.  This is explained in detail in \cite{fssy}, the basic idea is to split the factor of $\sinh k(\pi-\theta)$ into a sum of two exponentials, and then close the contour in the upper half plane for one term and in the lower half plane for the other term.  The contour rotations are always justified for real X.  The asymptotics at large $\omega$ can then be read off from the contributions of the various poles.  In fact all poles contribute terms that are subdominant to the bound state contribution.  The large $\omega$ dependence of this contribution can be read off from the continuation of the expression for $G_B(\theta)$, and it is linear in $\omega$.

Now recall that we started with one of the points fixed at $\theta=0$, so $\theta$ was just the geodesic distance on the unit three sphere.  After analytic continuation, this means that the Lorentzian two point function at general points in the geometry scales like the geodesic distance in the domain wall between the two points.  If we set them both to be at a particular X, choosing one to be at the pole of the two-sphere and the other to be at a fixed angle $\alpha$, then at large $\omega_1$, $\omega_2$ we find:\footnote{There is an interesting subtlety here in that there can be pairs of points in dS which have no geodesic connecting them.  Moreover at fixed $\alpha$ and large $\omega_1$, $\omega_2$, we invariably produce such pairs.  This is discussed in Appendix B, where we derive (\ref{scaling}).}
\be
\label{scaling}
\langle\chi(X,\omega_1,0) \chi(X,\omega_2,\alpha)\rangle \sim \omega_1 +\omega_2 + \log(\cos(1-\alpha))
\ee

The term $\omega_1 + \omega_2$ becomes infinite in the limit of large $\omega_1$, $\omega_2$, and corresponds to the infrared pileup of massless modes that leave the horizon as de Sitter time evolves.  The term $\log(\cos(1-\alpha))$ is finite in the same limit, and exhibits the long-range correlation characteristic of dynamical gravity.  By contrast note that in AdS, the correlator of a massless scalar falls off like $z^{2d}$ as $z\to 0$, and in an appropriate gauge so does the metric correlator.  Although this analysis describes massless scalar fluctuations of the field $\chi$, tensor fluctuations in the gauge
 $$
 \nabla^i   h_{mn}=  0
 $$
 $$
 h_m^m=0
 $$
 ($m,n$ run over the three directions in the surface D) are known to satisfy minimally coupled massless equations. In \cite{fssy} the details of calculating the tensor fluctuations were presented and one finds the same logarithmic behavior for the spatial components $h_{ij}$ that lie in 2-sphere $\Omega_2$.

 Thus it is clear that tensor perturbations of the domain wall D behave, in the infrared, as if 2+1 dimensional gravity were dynamical.

  \bigskip
  \bigskip
  \bigskip
  \bigskip

\setcounter{equation}{0}
\section{Three Limits}

 \subsection{Flat Ancestor Limit}

 Most likely there are no non-supersymmetric flat vacua except for those at the extreme limits of moduli space, where the  string coupling vanishes. However, it is possible to consider a  decay from a de Sitter vacuum with a very small cosmological constant, to an ADS vacuum with a  cosmological constant of much larger magnitude.  For many purposes it may be a good approximation to ignore the ancestor vacuum energy and treat it as flat. In this case the instanton geometry still satisfies \ref{2.1} but the spherical region is replaced by a portion of flat Euclidean space. Equation \ref{2.4} is unchanged but \ref{2.3} is replaced by
 \be
 a(x) =x.
 \ee
 In terms of the conformal coordinate $X$ the scale factor has the asymptotic form
 \be
 a \to e^X
 \ee
 as $X\to +\infty.$ Therefore the wave function \ref{2.12} becomes non-normalizable. In that case gravity decouples from the domain wall. One can therefore expect that when the cosmological constant in the ancestor is small enough, gravitational fluctuations in D become very weak.

 The problem with the decay from flat space is that it does not lead to an isolated crunching FRW geometry. The actual endpoint of the evolution will be much messier than this. What really happens is that an infinite number of bubbles nucleate and eventually coalesce in a grand crunch of the entire space. Therefore we expect the dual theory to be ambiguous and not have a UV completion. This seems right because UV complete theories are either asymptotically ADS or  asymptotically free, neither of which are dual to flat space.

\subsection{AdS/CFT Limit}
We can also consider the possibility that the ancestor vacuum is AdS.  The simplest thing to do is to have the AdS outside of the bubble be the same as the one inside, with the domain wall between carrying no energy.  There is no normalizeable zero mode, indeed this is just the ordinary AdS/CFT correspondence constructed on a dS slicing of the original AdS space.  The dual is a conformal field theory defined on background dS space, which can be precisely understood by defining it on the Euclidean $S^3$ and then continuing correlators.  The effective field theory on the domain wall is then just the product of integrating out the UV of the CFT, and is thus trivially UV complete.  This is illustrated in steps A-C above.

More interesting is to consider the possibility that the vacuum outside of the bubble is some other AdS', not equal to the one inside the bubble.  This situation was discussed extensively in \cite{daniel}: generically the existence of such a geometry corresponds to an instanton decay of AdS', and it leads to the nucleation of infinitely many bubbles that destroy any observer in AdS' in bounded proper time.  In fact the amount of proper time available to any particular observer was shown in \cite{daniel} to be independent of the decay rate, so the situation is even worse than the analogous problem in flat space just discussed.  In \cite{daniel}, this was interpreted as demonstrating that the idea of a metastable AdS is not precisely defined.

However this suggests a contradiction with the UV-complete situation discussed in the introduction, where a ``UV'' CFT living on a background dS space is deformed by a relevant operator in such a way that it flows to an ``IR'' CFT.  This RG flow corresponds to a bulk geometry with CdL symmetry that describes a bubble of ADS surrounded by an AdS', with a thick domain wall in between.  But shouldn't this allow for more bubbles to nucleate in AdS', leading to the pathology just discussed?  The answer must be no, since on the field theory side the construction is completely well-defined.

The way out of this was suggested in \cite{daniel} in a slightly different context: when the domain wall has thickness of order the AdS scale in the false vacuum, it is possible for the boundary conditions to forbid multi-bubble configurations.  To see how this works, note that in dS slicing the metric of $AdS_{d+1}$ is:
$$ds^2=dx^2+\sinh^2 x\left[-d\omega^2+\cosh^2 \omega d\Omega_{d-1}^2\right]$$
A massive scalar field in these coordinates will have $x\to \infty$ asymptotic behaviour:
$$\phi \to \alpha(\omega, \hat{\theta}) e^{-\Delta_+ x}+\beta(\omega, \hat{\theta}) e^{-\Delta_- x}$$
Here $\hat{\theta}$ represents the angular coordinates on $S^{d-1}$, and $\Delta_{\pm}=\frac{d}{2} \pm \sqrt{m^2+d^2/4}$ as usual in AdS/CFT.  If we wish to take the original CFT action and perturb by a relevant operator $\mathcal{O}$ that is dual to $\phi$, then we set boundary conditions where $\alpha$ is arbitrary but $\beta$ is equal to some constant $\beta_0$.  For the operator to be relevant, we must have $m^2<0$.  To see that this forbids multi-bubble configurations, we go back to global coordinates with metric:
$$ds^2= -(1+r^2)d\tau^2+\frac{dr^2}{1+r^2}+r^2d\Omega_{d-1}^2$$
The transformation between the two coordinates at large x is:
$$e^x=2 r \cos \tau$$
$$\tanh \omega = \sin \tau$$
This gives global boundary conditions for the field:
\be
\label{globalbc}
\phi \to \frac{\tilde{\alpha}(\tau,\hat{\theta})}{r^{\Delta_+}}+\frac{\beta_0}{(2\cos \tau)^{\Delta_-}} \frac{1}{r^{\Delta_-}}
\ee
Here $\tilde{\alpha}$ is an arbitrary function simply related to $\alpha$.  These boundary conditions would typically be interpreted as deforming the action by a relevant operator with a time-dependent coefficient.  It is easy to check that they break the full AdS symmetry down to the CDL symmetry, and that the broken symmetry is the one that would have allowed us to translate the initial bubble configuration around on the surface $\tau=0$.  Indeed for large $r$ the relevant isometry group has:
$$r'=r(\cosh \eta+\cos \tau \sinh \eta)$$
$$r' \cos \tau'=r(\cosh \eta \cos \tau+\sinh \eta)$$
Here $\eta$ is the boost angle of the transformation in the usual embedding geometry.  This clearly does not preserve the boundary condition (\ref{globalbc}).  What this means is that the boundary conditions only allow a bubble to be present in the \textit{center} of the geometry; there can be no bubbles nucleated anywhere else.  In fact these boundary conditions \textit{demand} the existence of a bubble: pure AdS with a constant $\phi$ does not respect them.\footnote{In \cite{daniel}, the boundary conditions that were considered were those first introduced in \cite{Hertog:2004rz}.  These corresponded to deforming the action by a multitrace operator, and preserved the full AdS symmetry.  As explained in \cite{daniel}, they allow either zero or one bubble to be present, but not more than one.}

Since there is only one bubble present, it is possible for an observer to outrun it for an arbitrarily long time without using an amount of energy that would destroy the asymptotic geometry, and thus we expect (and find) a precise dual description.  

We should emphasize that the trick of forbidding multiple bubbles by using the boundary conditions only worked because we were able to make the domain wall thick in false vacuum AdS units.  We can't make the case of a flat false vacuum precise in the same way since there is no scale that the domain wall can be thick with respect to.  

\subsection{Flat Bubble Limit}

Of particular interest is the limit in which the \cc \ in the bubble tends to zero (we now return to having a dS false vacuum). In that case the FRW region becomes a hat-geometry.  In this limit the dual description should become FRW/CFT duality \cite{hat}.  In the FRW/CFT duality, the field theory dual is a 2-dimensional Euclidean field theory coupled to a Liouville field with negative central charge. It is interesting to ask how this derives from the dual description of a crunch as the \cc \ tends to zero.

One very crude way to see the evolution from the crunch-dual to the hat-dual is to note that the cutoff distance on the hyperboloid D is given by the ADS radius. If we define the following length scales,
\begin{itemize}
\item $l_{ads } \ = \ $ the ADS radius of curvature in the bubble
\item $l_D \ = \ $  the radius of curvature  of the domain wall  de Sitter space
\item $l_{cutoff} \ = \ $ the cutoff length scale on the domain wall
\end{itemize}
Then
\be
{l_{cutoff}  \over l_D } \ = \ {l_{ads}  \over l_D }
\ee

If we keep the parameters of the ancestor and the domain wall fixed and let $l_{ads}$ grow, the cutoff length diverges relative to the domain wall radius. Obviously this is a very bad limit from the dual field theory point of view. The density of  ``lattice points" on D goes to zero.

This does not mean that the degrees of freedom disappear. First of all, the $N   \times N$ matrices grow as the density of lattice points decreases. Moreover there is an exponentially large volume of space near the future boundaries of D. Thus one can see that the degrees of freedom migrate to the surface $\Sigma.$ We will return to this in the next subsection.
 \begin{figure}[h]
\begin{center}
\includegraphics[width=4cm]{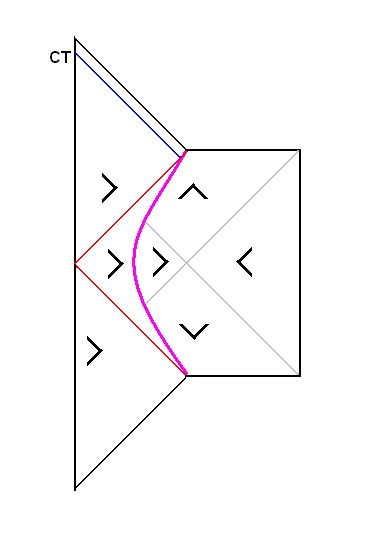}
\caption{Buosso-Penrose diagram for the hat geometry.}
\label{bpd}
\end{center}
\end{figure}
If the FRW/CFT duality is correct, how precise should it be? One may hope to answer this by asking how precise observations of a census taker \cite{hat} can be. According to the discussion at the beginning of this paper, a figure-of-merit would be the entropy bound on the census taker's past light cone. In Figure \ref{bpd} a Buosso-Penrose diagram is shown for the \cdl \ geometry for the case of vanishing \cc \ in the bubble. We also show a census taker at late time and his past light-cone. It is obvious that as the census taker tends to the tip of the hat, the entropy through the past light cone becomes infinite. Thus the precision of the census taker's observations is not limited by entropy bounds. For that reason the dual FRW/CFT description should be UV complete. In other words the Liouville-QFT model should not break down in the UV\footnote{One should also note that the divergent entropy bound is not affected by bubble collisions The effects of bubble collisions on FRW/CFT will be  the subject of a soon-to-be-published paper. Here we will just remark that bubble collisions induce a flow between fixed points.}.

 \subsection{FRW/CFT}

Let us return to the case of a crunching bubble dual to quantum gravity on 2+1 dimensional de Sitter space.

From the holographic principle, we know quantum gravity cannot be a a bulk quantum field theory in any  ordinary sense. Treating space as if there was a degree of freedom in each small volume is a blatant over-counting. This is true in de Sitter space as well as in flat and anti de Sitter space. The dual to a \cdl-nucleated crunching bubble must be something more subtle than cutoff quantum field theory on  2+1 dimensional de Sitter space.  dS/CFT \cite{ds-cft} is an approach that undoubtedly has some validity. That suggests if an ADS-crunching bubble is really dual to quantum gravity on D then we should be able to explore it using a 2+1 dimensional version of dS/CFT.

 The CFT of the dS/CFT duality lives on the future boundary of the (2+1 dimensional) de Sitter space, in this case the future boundary of D. It is
conformally a two sphere, and will be labeled\footnote{ We have intentionally labeled using the convention of \cite{hat}. The difference is that in \cite{hat}, the surface $\Sigma$ defined the intersection of the hat with the bulk ancestor, whereas here, it is the intersection of the crunch with ancestor.}
$\Sigma.$  According to dS/CFT there is a 2-dimensional Euclidean field theory living on $\Sigma$ that describes $D$.

Consider  a global space-like slice through D. Like $\Sigma$ the  slice has the topology of a 2-sphere. On that slice we specify a 2-dimensional Euclidean geometry $g_{ij}$. The assumption of \cite{ds-cft}, is that the Wheeler DeWitt wave function $\Psi(g)$ is given by the partition function for a 2-dimensional Euclidean CFT on a sphere with metric $g_{ij}.$ Without specifying the details, let us call the fields of that CFT $A(x)$ where $x$ labels a point on the sphere. (Note that $A$ does not include the metric $g$.)
 \be
 \Psi(g) = \int DA e^{-\int dx L(A;g)}.
 \ee
 In general the fields $A$ and the Lagrangian $L$ are complex and the field theory is not unitary.

 We are interested in expectation values of functionals of the metric $g$.
$$
 \langle F(g) \rangle = \int Dg  \
 \Psi^{\ast }(g)  F(g) \Psi(g)
 $$
 Writing $ \Psi^{\ast }(g)$ in the form
 \be
 \Psi^{\ast }(g) = \int DB^\ast e^{-\int dx L^\ast(B^\ast ;g)}
 \ee
 we may rewrite $ \langle F(g) \rangle $ in the form
 \be
 \langle F(g) \rangle = \int F(g) \  Dg  \ \int DA \ e^{-\int dx L(A;g)}  \int DB^\ast \ e^{-\int dx L^\ast(B^\ast ;g)}
\ee

There is nothing to prevent us from combining the integrations into a single integral over fields $A, \ B^\ast,$ and $g$.
 \be
 \langle F(g) \rangle = \int  Dg \ DA \ DB^\ast \  F(g) \  e^{-\int dx \{ L(A;g)+ L^\ast(B^\ast ;g)\}}
 \ee
In other words expectation values of $g$ are computed from a path integral  for a  system of 2-dimensional conformal fields coupled to a fluctuating metric\footnote{Silverstein and collaborators \cite{dsds}  have also proposed a dual description of de Sitter space which makes use of two conformal systems coupled to gravity. On the face of it their setup is different than ours. For example their dual is a Minkowski signature field theory while ours is Euclidean. However we suspect they the two descriptions  are related by analytic continuation.   }.

It is also obvious that the integration over the short distance degrees of freedom of $A$ and $B^\ast$ will generate a kinetic term for the metric. In fact, if the theory is written in conformal gauge, it will become a Liouville theory with a central charge that cancels the central charge of the matter fields, $A,B$. The central charge of $A,B$  is proportional to the original ancestor de Sitter entropy which in the semiclassical regime is large. Therefore the central charge of the Liouville field will be large and \it negative. \rm  Liouville theories with negative central charge are called time-like. The resulting dS/CFT theory is a
time-like  Liouville theory coupled to a large number of matter fields.

We expect that the correlation functions produced by this construction will tend to the same correlators that are computed by the FSSY procedure  in the FRW region, and then pulled towards the boundary from within the bubble.  So one might think that this means this geometry would admit an FRW/CFT type dual\cite{hat}.\footnote{Note that the Lorentzian correlators  in the FRW region, and correlators near the domain wall, are  obtained by analytic continuation of the same Euclidean correlators.  Recall that the continuation from $\theta$ to $\omega$ or $R$ is linear in both cases and agrees for large values, so the two sets of correlators are smoothly related at large $R$ or $\omega$.  Another way to say this is that the geometry is completely smooth on the null surface separating the two regions.  So if we take points and pull them towards the boundary two-sphere $\Sigma$, the limit should be the same from both sides. }
If true this would lead to a puzzle because the ADS-crunch geometry embedded in a de Sitter ancestor, should not have a completely precise description. There must be some reason why the FRW/CFT dual is  UV complete for the hat, but not for the ADS-crunch.
In order to explain how FRW/CFT may fail to be UV complete we will review the basic ideas of \cite{hat}.

First begin with 3-dimensional Euclidean ADS/CFT. The metric of ADS is
\bea
ds^2\eq   a^2 [dR^2 + \sinh^2{R} \ d\Omega_2^2]\cr
\eq a^2 d\h3^2
\label{2.21}
\eea
where $a$ is the radius of curvature and $\h3$ represents the 3-dimensional hyperbolic plane. The dual CFT lives on the unit 2-sphere $\Omega_2$ with no $R$ coordinate.

For large $R$ \ref{2.21} becomes
\be
ds^2=   a^2 [dR^2 + {1\over 4}e^{2{R}} \ d\Omega_2^2]
\label{2.21p}
\ee

The $R$ coordinate in the bulk space is represented on the CFT side as the RG reference scale. More exactly it is the logarithm of the reference momentum scale. To define a renormalized action at scale $R$ one integrates out the high momentum modes down to momentum $e^R$. This defines a renormalized theory at the reference scale $e^R$. Note that the renormalized action depends on the reference scale, but in just such a way that physical RG invariant quantities do not.

Another way to think about it is to consider the metric on the cutoff surface at radial coordinate $R$. From \ref{2.21p} we see that apart from a factor of $a^2/4$, the metric scales like
\be
ds^2 \approx e^{2{R}} \ d\Omega_2^2
\label{2.21pp}
\ee
Instead of thinking of RG flow in terms of variation with respect to a reference momentum scale, we can think of variation with respect to a reference metric.

One can also start with an action at reference scale $R$ and try to integrate the RG equations backward to obtain an effective action at $R'\ > \ R$. Sometimes this is possible and the low energy theory is said to have  a UV completion. Sometimes it is not possible---a Landau pole appears---as  explained in Section 1.

Now let us turn to FRW/CFT.
There are now two coordinates that have to emerge from RG flow: $R,$ and  time.
 Recall that in \cite{hat} the Liouville field was identified with scale-factor time in the FRW geometry. To see why, consider the metric in the bubble. Using conformal time
\bea
ds^2 \eq a(T)^2 [ -dT^2 +  d\h3^2]\cr
\eq a(T)^2 [ -dT^2 + dR^2 + \sinh^2R d\Omega_2^2]
\label{2.21ppp}
\eea
where for the hat-geometry, at  large $T,$ $a(T) $ is given by
\be
a(T) \approx e^T
\ee

Now consider the metric on the 2-sphere $\omega_2$ for large $R$ and $T$.
\bea
ds^2 \eq  a(T)^2e^{2R} d\Omega_2^2 \cr
&\approx& e^{2T}e^{2R} d\Omega_2^2
\eea
We see that the metric is the product of two factors: a reference metric as in \ref{2.21pp}, and a second Liouville factor $e^{2\phi}$ with $\phi = T$.

In a Liouville theory the RG flow can be considered to be 2-dimensional in the sense that one can vary with respect to the reference metric and with  respect to the full metric. We can also think of there being two renormalization scales, $e^R$ and a ``fishnet" scale $e^{R+T}$  \cite{hat}.

The theory at scale $e^{R+T}$ can be thought of as the bare lattice theory. Integrating out the degrees of freedom between $e^{R+T}$ and $e^R$ gives an effective theory on the reference geometry. The question of the UV completeness of FRW/CFT is the question of whether the RG flow can be run backward all the ways to $T=\infty$ without obstruction.

Now consider the behavior of the scale factor $a(T)$ in the ADS-crunching  geometry.  Instead of monotonically increasing, it reaches a maximum, and then decreases to zero at the singular crunch. From the point of view of the Liouville theory, something happens to the RG flow when the ratio of the cutoff momentum to the reference momentum reaches a certain value. Evidently the RG flow cannot be pushed past that point.  Since there is no observer who can study the dual theory at scales shorter than this, the most likely conclusion is that the theory fails to be UV-complete beyond this point.  Certainly there is no doubt that this turnaround cannot be described by a single conventional RG flow and that the standard FRW/CFT interpretation becomes untenable.

By contrast if the \cc in the bubble vanishes, the scale factor is monotonic. In that case $a(T) \sim e^T$ for large time. This is the case of a hat-geometry for which  FRW/CFT was originally designed.
Thus we suspect that for decays from dS a true dual can only exist for a hat and not for an ADS crunch.

\bigskip
\setcounter{equation}{0}

\section{A Conjecture}

It is very interesting that Maldacena's proposal relates two systems, both of which seem to resist exact quantum descriptions, namely crunches and de Sitter space. The question is, what general lessons can we derive from such examples, and from more precise dualities like ADS/CFT, Matrix Theory, and FRW/CFT?

The reason that de Sitter space and ADS-crunches   (those that nucleate from a \cdl \ bubble) are imprecise is that in both geometries the entropy (information) that can be gathered by any Census Taker is finite and bounded by the Bousso bound.
If we assume that the dual descriptions of these geometries are no more precise than the thing they are describing, then the duals should also be imprecise.

In a given geometry different Census Takers may have different entropy bounds on their past light-cones. That raises the question of which Census Taker's bound controls the maximum precision of the dual description? The most reasonable answer is the
Maximal Census Taker, i.e., the one who's past light cone has the maximal entropy bound. Thus we propose the following conjecture:

  \bigskip
  \bigskip

\it The maximum precision of a dual description of a cosmological geometry is determined by the entropy bound of the Maximal Census Taker. The description can only be exact---ultraviolet complete---if there exists a Census Taker with an infinite entropy bound.  \rm

\bigskip

\bigskip

 \begin{figure}[h]
\begin{center}
\includegraphics[width=4cm]{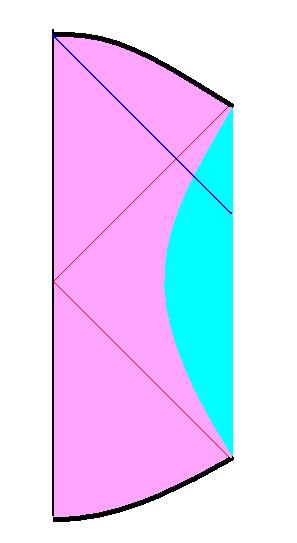}
\caption{Penrose diagram of a UV-complete crunch.  The ``UV'' CFT is shown in blue and the ``IR'' CFT in pink.  The backward lightcone of an observer who hits the crunch is the blue line.}
\label{crunchpd}
\end{center}
\end{figure}
This conjecture is consistent with all of the examples we have discussed here, as well as other well-known examples involving black holes.  We believe that it is a good necessary condition for the existence of a precise dual.

There is however a final subtlety to discuss in determining whether or not it is a sufficient condition.  It is possible for an observer who falls into a singularity to have a past lightcone to which we can assign an infinite entropy bound.  For example consider an observer in the ``UV-complete crunch geometry'' described above, and depicted here in figure \ref{crunchpd}.  The observer who follows a timelike geodesic from one crunch to the other in the center of the geometry will have a backwards lightcone that reaches the AdS boundary and thus will be able to see an infinite amount of entropy.  Such an observer will not really have time to process all of that information, so it seems that the existence of such observers is not quite enough to demand a precise dual according to the philosophy of our introduction.  

The resolution in this example, as mentioned above, is that by using rockets observers can accelerate away from the domain wall and thus survive for arbitrarily long amounts of time before falling into the final crunch.\footnote{Indeed this geometry is very similar to the eternal AdS black hole, where it is also true that observers will need to use rockets to avoid falling in to the singularity.}  It is these observers who would really be able to probe their environment with the precision a continuum dual theory would provide.  This subtlety is only present when there is a spacelike singularity in the future, and we expect that in all such cases, if there exists a point on spacelike future infinity whose backwards lightcone can have an infinite entropy bound, then there exist timelike curves, attainable using rockets, that have arbitrarily long proper time AND which reach future infinity at points whose backwards lightcones have infinite entropy bounds.  This is certainly true for all examples we have considered here.  A heuristic argument for it being true in general is that the presence of an infinite entropy bound requires the Penrose diagram to have a boundary with infinite radius, and this boundary must join with the future spacelike singularity at some point.  Observers can prolong their lives by accelerating towards that point.  This argument is admittedly crude, and it would be interesting to study this more systematically in general relativity.

\bigskip
\bf Acknowledgments: \rm

We are very grateful to Eva Silverstein for a number of important insights about the existence of gravity on the domain wall, and about the possibility of ultraviolet complete models of crunches.

We'd also like Simeon Hellerman, Edgar Shaghoulian, Steve Shenker, Yasuhiro Sekino, and Douglas Stanford for helpful remarks.  This work was supported in part by NSF grant PHY-0756174.

\appendix

\section{Thin-wall Formulas}
We record here explicit formulas for the two point function in the thin-wall approximation to the geometry.

Using our thin-wall metric (\ref{metric}), we find:
 \be
 U(X)=\alpha\delta(X) +\theta(-X)\left(1+\frac{2}{\sinh^2(X-X_0)}\right)+\theta(X)\left(1-\frac{2}{\cosh^2(X+X_1)}\right)
 \ee
 The coefficient of the delta function, $\alpha,$ was defined in equation (\ref{tension}).
This potential and its bound state are shown in figure 3 for a typical choice of the parameters.

The scattering wave functions for $X<0$, $k>0$ are given by:
\be
u_k(X)=\frac{1}{1-\frac{i}{k}}e^{ikX}\left(1+\frac{i}{k}\coth(X-X_0)\right)+\mathcal{R}(k)\frac{1}{1+\frac{i}{k}}e^{-ikX}\left(1-\frac{i}{k}\coth(X-X_0)\right)
\ee
The reflection coefficient can be found using another supersymmetric quantum mechanics trick of FSSY:
\be
\mathcal{R}(k)=\frac{k+i}{k-i} \frac{\alpha}{\alpha-2ik}
\ee
For $X<0$, $k<0$, we have:
\be
u_k=\mathcal{T}(-k)\frac{1}{1-\frac{i}{k}}e^{ikX}\left(1+\frac{i}{k}\coth(X-X_0)\right)
\ee
The transmission coefficient is given by $\mathcal{T}(k)=1+\mathcal{R}(k)$.  Combining these equation gives the correlator as:
$$
\hat{G}(X_1,X_2,\theta)=\int_{-\infty}^{\infty} \frac{dk}{2\pi}G_k(\theta)\left[\frac{1}{1+k^2}e^{ik(X_2-X_1)}(k-i \coth(X_1-X_0))(k+i \coth(X_2-X_0))\right.
$$
$$
+\left.\frac{\mathcal{R}(k)}{(k+i)^2}e^{-ik(X_1+X_2)}(k-i \coth(X_1-X_0))(k-i \coth(X_2-X_0))\right]
$$
\be
\label{2pointthin}
+u_B(X_1)u_B(X_2) G_B(\theta)
\ee

One can check that these formulae reduces to the FSSY expressions in the limit $X_0\to -\infty$, which cooresponds to taking $l_{ads}\to \infty$ while fixing the domain wall tension.  The first term in the integral is the piece that would be present if there was no domain wall and the AdS extended all the way out to its boundary.

\begin{figure}[t!]
\centering
{\includegraphics[scale=1]{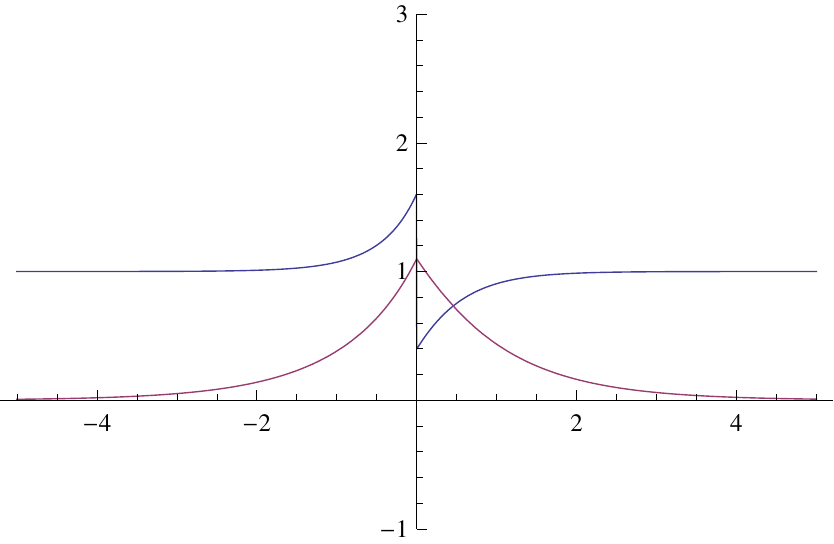}}
\caption{A plot of the potential U(X) in blue and the bound state wave function a(X) in pink. There is also a delta-function at the origin for the potential, which is not shown.}
\end{figure}

\section{Geodesic Distance in de Sitter space}
We found that the Euclidean two point function of a massless scalar on the CDL geometry was proportional to the geodesic distance connecting the two points on the three sphere, with metric:
$$ds^2=d\theta^2+\sin^2\theta\left(d\phi^2+\sin^2 \phi d\psi^2\right)$$

Since we will continue in $\theta$, it is convenient to have a formula for the distance between points at a general $\theta_1$, $\theta_2$.  We can choose both points to have $\psi=0$ by doing rotations, and we can choose one to have $\phi=0$ and the other to have $\phi=\alpha$.  By embedding the sphere in flat space, we can take the inner product of the unit vectors pointing towards each of them to find:
$$\cos \theta = \cos\theta_1 \cos \theta_2+\sin \theta_1\sin \theta_2 \cos \alpha$$
Here $\theta$ is the angle between the two points on the sphere, and also the geodesic distance so it is what the Euclidean correlator is proportional to.  If we do the continuation (\ref{continuation}), we find:
$$\cos \theta = -\sinh \omega_1 \sinh \omega_2+\cosh \omega_1 \cosh \omega_2\cos\alpha$$
When the right-hand side is less than one in absolute value, there is a spacelike geodesic connecting the two points and $\theta$ is real.  But for large $\omega_1$, $\omega_2$, the right hand side goes to $e^{\omega_1+\omega_2}( \cos \alpha-1)$ which becomes large in absolute value.

We can make use of the fact that for large negative x:
$$\arccos x \to \pi-i\log(-x)$$
To see that for large $\omega_1$ and $\omega_2$, we have:
$$\theta \to \pi+i\left(\omega_1+\omega_2 +\log(1-\cos \alpha)\right)$$
Since the real part of $\theta$ goes to $\pi$ in this limit and the imaginary part becomes large, our asymptotic analysis in section 2.3 was justified, so we have finally shown (\ref{scaling}).

\end{document}